\documentclass[a4paper,fleqn,usenatbib]{mnras}

\usepackage[T1]{fontenc}
\usepackage{ae,aecompl}

\usepackage{graphicx}	% Including figure files
\usepackage{amsmath}	% Advanced maths commands
\usepackage{amssymb}	% Extra maths symbols

\title[Optical observations of supernova iPTF15dld]{Optical photometry and spectroscopy of the low-luminosity, broad-lined Ic supernova iPTF15dld\thanks{Based on observations made with the {\it Copernico} telescope (Asiago, Italy) of  INAF - Osservatorio Astronomico di Padova;
with the NTT at the European Organisation for Astronomical Research in the Southern Hemisphere, Chile, as part of  the Public ESO Spectroscopic Survey for Transient Objects Survey (PESSTO) ESO programs 188.D-3003, 191.D-0935; with the Telescopio Nazionale Galileo, operated  by the Fundaci\'on Galileo Galilei of INAF, with the Liverpool Telescope, operated  by Liverpool John Moores University  with financial support from the UK Science and Technology Facilities Council, and with the Nordic Optical Telescope, operated by the
Nordic Optical Telescope Scientific Association, all three on the island of La Palma at the Spanish
Observatorio del Roque de los Muchachos of the Instituto de Astrof\'isica de Canarias.}}

\author[E. Pian et al.]{
E. Pian,$^{1,2}$\thanks{E-mail: elena.pian@sns.it}
L. Tomasella,$^{3}$
E. Cappellaro,$^{3}$
S. Benetti,$^{3}$
P. A. Mazzali,$^{4,5}$
\newauthor
C. Baltay,$^{6}$ 
M. Branchesi,$^{7,8}$
E. Brocato,$^{9}$
S. Campana,$^{10}$
C. Copperwheat,$^{4}$
\newauthor
S. Covino,$^{10}$
P. D'Avanzo,$^{10}$
N. Ellman,$^{6}$
A. Grado,$^{11}$
A. Melandri,$^{10}$
\newauthor
E. Palazzi,$^{1}$
A. Piascik,$^{4}$
S. Piranomonte,$^{9}$
D. Rabinowitz,$^{6}$
G. Raimondo,$^{12}$
\newauthor
S. J. Smartt,$^{13}$
I.A. Steele,$^{4}$
M.  Stritzinger,$^{14}$
S. Yang,$^{3}$
S. Ascenzi,$^{9}$ 
\newauthor
M. Della Valle,$^{11,15}$
A. Gal-Yam,$^{16}$
F. Getman,$^{11}$ 
G. Greco,$^{7,8}$
C. Inserra,$^{13}$
\newauthor
E. Kankare,$^{13}$
L. Limatola,$^{11}$
L. Nicastro,$^{1}$
A. Pastorello,$^{3}$ 
L. Pulone,$^{9}$
\newauthor
A. Stamerra,$^{2,17}$
L. Stella,$^{9}$
G. Stratta,$^{7,8}$
L. Tartaglia,$^{3}$
and M. Turatto$^{3}$ 
\\
$^{1}$ INAF, Istituto di Astrofisica Spaziale e Fisica Cosmica di Bologna, Via Gobetti 101, 40129 Bologna, Italy\\
$^{2}$ Scuola Normale Superiore, Piazza dei Cavalieri 7, 56126 Pisa, Italy\\
$^{3}$ INAF, Osservatorio Astronomico di Padova, Vicolo dell'Osservatorio 5, 35122 Padova, Italy\\
$^{4}$ Astrophysics Research Institute, Liverpool John Moores University, IC2, Liverpool Science Park, 146 Brownlow Hill, Liverpool L3 5RF, UK\\
$^{5}$ Max-Planck-Institut f\"ur Astrophysik, Karl-Schwarzschild-Str. 1, D-85748 Garching, Germany\\
$^{6}$ Physics Department, Yale University, P.O. Box 208120, New Haven, CT 06520, USA\\
$^{7}$ Universit\'a degli Studi di Urbino `Carlo Bo', Dipartimento di Scienze Pure e Applicate, Piazza della Repubblica 13, I-61029, Urbino, Italy\\
$^{8}$ INFN, Sezione di Firenze, I-50019 Sesto Fiorentino, Firenze, Italy\\
$^{9}$  INAF, Osservatorio Astronomico di Roma, Via di Frascati, 33, I-00040 Monteporzio Catone, Italy\\
$^{10}$ INAF, Osservatorio Astronomico di Brera, Via E. Bianchi 46, 23807 Merate (LC), Italy\\
$^{11}$ INAF, Osservatorio Astronomico di Capodimonte, salita Moiariello 16, 80131, Napoli, Italy\\
$^{12}$  INAF, Osservatorio Astronomico di Teramo, Via M. Maggini s.n.c., 64100 Teramo, Italy\\
$^{13}$  Astrophysics Research Centre, School of Mathematics and Physics, Queen's University Belfast, Belfast BT7 1NN, UK\\
$^{14}$ Department of Physics and Astronomy,  Aarhus University, Ny Munkegade 120, 8000, Aarhus C, Denmark\\
$^{15}$ International Center for Relativistic Astrophysics, Piazza delle Repubblica, 10, 65122-Pescara, Italy\\
$^{16}$ Department of Particle Physics and Astrophysics, Weizmann Institute of Science, Rehovot 76100, Israel\\
$^{17}$  INAF, Osservatorio Astronomico di Torino, Via Osservatorio 30, 10025 Torino, Italy
}

\date{Accepted XXX. Received YYY; in original form ZZZ}

\pubyear{2016}

\begin{document}
\label{firstpage}
\pagerange{\pageref{firstpage}--\pageref{lastpage}}
\maketitle

\begin{abstract}
Core-collapse stripped-envelope supernova (SN) explosions reflect the diversity of  physical parameters and evolutionary paths of their massive star progenitors. 
We have observed the type Ic  SN iPTF15dld ($z = 0.047$), reported by the Palomar Transient Factory. Spectra were taken starting 20 rest-frame days after maximum luminosity and are affected by a young stellar population background.  Broad spectral absorption lines associated with the SN are detected over the continuum, similar to those measured for broad-lined, highly energetic SNe Ic.  The light curve and maximum luminosity are instead more similar to those of low luminosity, narrow-lined Ic SNe.  This suggests a behavior whereby certain highly-stripped-envelope SNe do not produce a large amount of $^{56}$Ni, but the explosion is sufficiently energetic that a large fraction of the ejecta is accelerated to higher-than-usual velocities.    We estimate SN iPTF15dld had a main sequence progenitor of 20-25 M$_\odot$, produced a $^{56}$Ni mass of  $\sim$0.1-0.2 M$_\odot$, had an ejecta mass of [2-10] M$_\odot$, and a kinetic energy of [1-18]  $\times10^{51}$ erg.
\end{abstract}


%%%%%%    INTRODUCTION  %%%%%%%

\section{Introduction}

Stripped-envelope supernovae (SNe), i.e. core-collapse SNe that have  lost their hydrogen envelope, and  retained  (type Ib) or lost (type Ic) their helium envelope,  are the progeny of massive stars \citep{Nomoto1988,Heger2003}.  Their  light curves  \citep{Brown2009,Drout2011,Li2011,Bianco2014,Pritchard2014,Taddia2015,Lyman2016,Prentice2016}
and spectra \citep{Filippenko1997,Matheson2001,Modjaz2014}  display significant diversity,
owing to the many different parameters of the exploding stellar cores (masses, rotation rates, metallicity, multiplicity), and possibly to the different degree of asphericity of the explosion \citep{Wheeler2000}.

Type Ic SNe  characterized by broad  absorption lines or high photospheric velocities ($\sim$15000-20000 km s$^{-1}$ at maximum luminosity), and hence high kinetic energies ($\sim 10^{52}$ erg), accompany the majority of long-duration gamma-ray bursts (GRBs; Woosley \& Bloom 2006).  This  points  to the presence of an extra  source of energy, besides radioactive $^{56}$Ni, i.e. a rotating, and possibly accreting, inner compact remnant.   This ``engine"  may play a role also in $\sim$5\% of all detected SNe Ic with high photospheric velocities, that are however  not accompanied by GRBs \citep{Mazzali2002,Valenti2008a,Soderberg2010,Corsi2011,Pignata2011}.

This heterogeneous phenomenology needs to be mapped  onto the properties of the progenitors and the explosions, and the intrinsic physical effects must be distinguished from those generated by differences in  the viewing angle toward the explosion symmetry axis.  Therefore, it is important to observe these SNe accurately and to build a complete physical scenario.
Optical multi-colour searches with very large field-of-view cameras and high cadence are ideal to detect a large number of core-collapse SNe, that are rather common, but often faint and buried in their host galaxy's starlight.
During the wide field optical searches of the huge sky localization uncertainty area of the gravitational wave 
candidate detected by the Advanced LIGO interferometers (aLIGO; Abbott et al. 2016) on 22 October 2015 \citep[called G194575, ][]{lvc2015} and subsequently flagged  as a low probability event  \citep[false alarm rate of 1/1.5 per days, ][]{lvc2016},  many multi-wavelength transients  were detected that are unrelated with the event (see Corsi et al. 2016; Palliyaguru et al. 2016, and references therein), a fraction of which were spectroscopically classified.   Among these is   iPTF15dld.

SN iPTF15dld was detected \citep{Singer2015}  by the 48inch Oschin telescope at Mount Palomar during the  intermediate Palomar Transient Factory (PTF) survey \citep{Rau2009,Law2009,Kulkarni2013} on 23 October, 08:15 UT
at coordinates RA = 00:58:13.28, Dec =   -03:39:50.3 with a magnitude of 18.50  \citep[Mould $R$ filter, AB system, ][]{Ofek2012}.  The initial identification as a Seyfert 2 galaxy at $z = 0.046$ \citep{Tomasella2015a}, based on a preliminary  spectral analysis,  was later revised to the classification as a broad-lined type Ic SN \citep{Benetti2015}.    We here very slightly revise the redshift to $z = 0.047$ based on accurate analysis of the host galaxy emission lines.  This corresponds to a distance of 200 Mpc using $H_0 = 73$ km~s$^{-1}$~Mpc$^{-1}$ \citep{Riess2016}, and a flat cosmology with $\Omega_m = 0.31$   \citep{PlanckColl2015}.   The Galactic extinction along the SN line of sight is $A_V = 0.085$ mag \citep{SchlaflyFink2011}.
The SN was also independently discovered as LSQ15bfp on 5 October 2015 with  $V = 19.5$ mag during the La Silla QUEST survey   \citep[LSQ, ][]{Baltay2013,Walker2015}  by  \citet{Rabinowitz2015} who also report a 
pre-discovery detection on 3 October 2015  at $V = 20.2$ mag and a brightening of 0.7 mag in 2 days suggesting that this date must be very close to  explosion time.  The object was also detected by Pan-STARRS as PS15crl in 6 separate 
exposures on 23 October 2015  (see Smartt et al. 2016, and Huber et al. 2015 for a description of the current Pan-STARRS surveys\footnote{http://star.pst.qub.ac.uk/ps1threepi/}).  The Pan-STARRS reference images show a very blue  starburst region that is superimposed on a  larger spiral galaxy.
\citet{Corsi2016}, who reported early optical photometry and a spectrum on 7 November 2015, detected no significant X-ray or radio emission for this SN (see also Evans et al. 2015; 2016).

Here we present the {\it Swift}/UVOT and ground-based optical observations of the  SN, including those preliminarily reported in \citet{Tomasella2015b} and \citet{Steele2015}, and  additional spectra  acquired within the PESSTO program \citep{Smartt2015}.
We adopt 3 October 2015 as the date of explosion, with an uncertainty of one day. 

%------------------ Figure 1: Asiago gri-coadded image ---------------
\begin{figure*}
\includegraphics[width=0.9\textwidth]{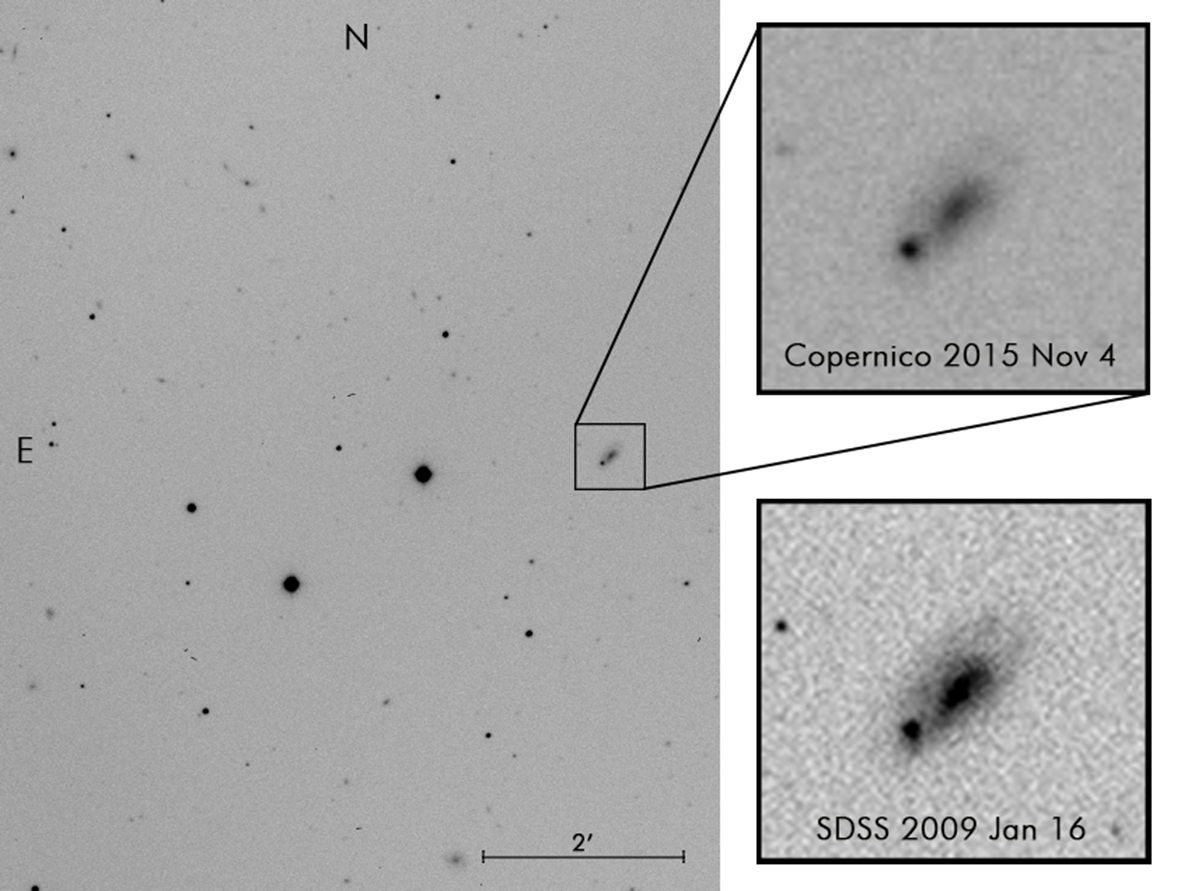}
\caption{Images of the field of iPTF15dld in  $r$-band  (exposure time of 120 seconds)  taken on 4 November 2015 with the  1.82m {\it Copernico} telescope (larger panel on the left and enlargement centered on the host galaxy on the top-right smaller panel) and from the SDSS prior to explosion (smaller bottom-right panel, covering the same area as the small top-right panel).}
\label{fig:snfield}
\end{figure*}
%---------------------------------------------------------------------------------------

%%%      SECTION 2:  OBSERVATIONS AND DATA ANALYSIS    %%%

\section{Observations and Data Analysis}
\label{data:obs}

Optical photometry and spectroscopy of the SN were acquired at the  1.82m {\it Copernico} telescope at Cima Ekar (Asiago, Italy),  at the Telescopio Nazionale Galileo (TNG), Nordic Optical Telescope (NOT) and Liverpool Telescope (LT, Steele et al. 2004) at the Canary Islands (Spain), at the ESO NTT and 1m Schmidt  telescope as part of the PESSTO and LSQ surveys, respectively.  UV photometry was taken with the UVOT instrument onboard the  {\it Swift} satellite.
%, and at the two 1.8m telescopes in Haleakala (Hawaii, USA) as part of the Pan-STARRS  survey.   
The logs of optical photometric and spectroscopic observations are  reported in  Tables \ref{tab:gb_obs_phot} and \ref{tab:gb_obs_spec}, respectively.    The exposure times were typically 5-10 minutes for the photometry and 20-40 min for spectroscopy.  These data were reduced following standard tasks within the IRAF\footnote{IRAF is distributed by the National Optical Astronomy Observatory, which is operated by the Association of Universities for Research in Astronomy (AURA) under a cooperative agreement with the National Science Foundation.}  reduction package.

\subsection{Photometry}
\label{data:photred}

The  $r$-band image of the SN field obtained at the   {\it Copernico}  telescope   is presented in   Figure~\ref{fig:snfield}.  The SN exploded in the outskirts of a spiral galaxy, in a starburst region that is marginally resolved both in our and in the SDSS images ($\sim 2.5^{\prime\prime}$ angular size) and contaminates dramatically the  measurements of the SN  in the bluer bands (see Sect. \ref{res:hostgalaxy}).  

Given the complex  background, the SN magnitudes were measured via  template subtraction.
For this purpose we used the SNOoPY package\footnote{SNOoPy: a package for SN photometry, http://sngroup.oapd.inaf.it/snoopy.html} developed by one of us (E. Cappellaro): this is a collection of python scripts based on publicly available tools. In particular, for template subtraction we used the  ``hotpants"  package\footnote{http://www.astro.washington.edu/users/becker/v2.0/hotpants.html}.  
For the  LSQ observations  we used images of the field taken by  the LSQ in 2012 as subtraction templates; while for the {\it ugriz}  photometry we used SDSS images, which provide a solid estimate of the pre-explosion background.   
SN magnitudes in the template-subtracted images    were  measured  by PSF fitting. We found PSF fitting is less sensitive to  background fluctuations compared with standard aperture photometry.
The LSQ images are unfiltered, but close to the $r$ filter, therefore the magnitudes resulting from the photometry were converted to this  band using a calibrating sequence of field stars. 

Starting on 6.97 November 2015,   UT and ending on  7.43 November 2015, UT the  {\it Swift}  satellite  observed the target (see observing log in  Table~\ref{tab:uvotobs}).  The UVOT camera measurements in the optical and UV were reduced according to Brown et al. (2015) and calibrated following Poole et al. (2008) and Breeveld et al.  (2010).    Aperture photometry with a radius of   $5^{{\prime}{\prime}}$ with background estimated from a nearby sky area yielded  the magnitudes reported in  Table~\ref{tab:uvotobs}.   

% ----------  Table 1: Log of ground-based photometric observations  -----------
\begin{table*}
\centering
\caption{Ground-based photometry$^a$ of iPTF15dld.}
\begin{tabular}{ccccc}
\hline
\hline
MJD                  &  UT                          & Tel.+instr./Survey  & r & i  \\
\hline
57284.17 & 2015 Sep 19.17 & LSQ$^b$                  & $>18.8$               & ...                           \\
57298.29 & 2015 Oct 3.29  & LSQ                            & $20.2 \pm 0.4$ & ...                            \\
57300.20 & 2015 Oct 5.20  & LSQ                          & $19.0 \pm 0.4$     & ...                            \\
57306.17 & 2015 Oct 11.17  & LSQ                        & $18.4 \pm 0.4$   & ...                            \\
57312.16 & 2015 Oct 17.16  & LSQ                        & $18.4 \pm 0.3$   & ...                            \\
57318.17 & 2015 Oct 23.17  & LSQ                        & $19.4 \pm 0.5$   & ...                            \\
57318.98 & 2015 Oct 23.98 &  PS$^c$                   &  ...                           &  $18.80 \pm 0.04$  \\ 
57319.15 & 2015 Oct 24.15  & LSQ                        & $19.2 \pm 0.4$   & ...                            \\
57324.13 & 2015 Oct 29.13  & LSQ                       & $20.1 \pm 0.5$    & ...                            \\
57330.94 & 2015 Nov 4.94 & 1.82m+AFOSC   & $19.9 \pm 0.1$   & $19.9 \pm 0.2$    \\
57332.11 & 2015 Nov 6.11  & LSQ                         & $20.5 \pm 0.4$    & ...                             \\
57332.87 & 2015 Nov 6.87 & 1.82m+AFOSC   & $20.2 \pm  0.09$   & $20.5 \pm 0.2$    \\
57332.92 & 2015 Nov 6.92 & TNG+LRS           & $20.0 \pm 0.1$    & ...  \\
57333.85 & 2015 Nov 7.85 & 1.82m+AFOSC   & $19.9 \pm 0.2$   & $20.4 \pm 0.1$     \\
57334.10 & 2015 Nov 8.10 & LSQ                          & $20.6 \pm 0.4$   & ...                              \\
57334.87 & 2015 Nov 8.87 & 1.82m+AFOSC   & $20.0 \pm 0.2$   & $20.4 \pm 0.1$     \\
57338.84 & 2015 Nov 12.84 & 1.82m+AFOSC & $20.1 \pm 0.2$   & $20.7 \pm 0.3$      \\
57341.92 & 2015 Nov 15.92 & 1.82m+AFOSC & $20.3 \pm 0.2$   & $20.8 \pm 0.4$      \\
57342.85 & 2015 Nov 16.85 & 1.82m+AFOSC & $20.4 \pm 0.2$   & $20.8 \pm 0.2$      \\
57344.90 & 2015 Nov 18.90 & 1.82m+AFOSC & $20.5 \pm 0.2$   & $21.1 \pm 0.3$      \\
57358.82 & 2015 Dec 2.82 & 1.82m+AFOSC   & $20.6 \pm 0.2$   & $21.1 \pm 0.3$      \\
57361.83 & 2015 Dec 5.83 & 1.82m+AFOSC   & $20.8 \pm 0.3$   & $>21.1$                   \\
57363.83 & 2015 Dec 7.83 & 1.82m+AFOSC   & $20.5 \pm 0.3$   & $>20.7$                   \\
57366.77 & 2015 Dec 10.77 & 1.82m+AFOSC & $20.7 \pm 0.1$   & $21.5 \pm 0.2$      \\
57373.76 & 2015 Dec 17.76 & 1.82m+AFOSC & $21.0 \pm 0.2$   & $21.5 \pm 0.3$      \\
57374.72 & 2015 Dec 18.72 & 1.82m+AFOSC & $20.8 \pm 0.2$     & $>21.1$                   \\
57399.83 & 2016 Jan 12.83  & NOT+ALFOSC      & $21.1 \pm 0.1$   & $21.9 \pm 0.3$       \\ 
\hline
\hline 
\noalign{\smallskip}
\multicolumn{5}{l}{$^a$  The magnitudes are galaxy-subtracted and not corrected for Galactic extinction.} \\
\multicolumn{5}{l}{$^b$ The La Silla QUEST survey uses the 1m ESO Schmidt telescope at the La Silla}\\ 
\multicolumn{5}{l}{ ~~~ Observatory with the 10 square degree CCD camera.}\\
\multicolumn{5}{l}{$^c$ This value was reported in Rabinowitz et al. (2015) from the Pan-STARRS}\\
\multicolumn{5}{l}{ ~~~ Survey for Transients (Huber et al. 2015).}\\
\end{tabular}
\label{tab:gb_obs_phot}

\end{table*}
% ------------------------------     END OF TABLE 1     ----------------------------------

\subsection{Spectroscopy}
\label{data:specred}

After bias and flat-field correction, the SN spectra were extracted and wavelength-calibrated  through the use of  arc lamp spectra.  Flux calibration was derived from observations of spectrophotometric standard stars obtained, when possible, on the same night as the SN. Corrections for the telluric absorption bands were derived using telluric standards. In some cases, non-perfect removal can affect the SN features that overlap with the strongest atmospheric features, in particular with the telluric O2 A band at 7590-7650 \AA.

In order to subtract the starburst contribution from the SN spectra, we used  the template spectra of star-forming galaxies by \citet{Kinney1996}.   The best fitting template was chosen by matching the  colours of the  starburst region as measured on the pre-explosion SDSS images (Table \ref{tab:hostgalaxy}):  this indicated a preference for a template with moderate intrinsic absorption ($0.11 < E_{B-V} < 0.21$, Kinney et al. 1996), as independently indicated also by the UVOT detections in the UV filters.    The spectral template  was fitted with a low order polynomial (to reduce noise in subtraction);  the relative contributions of the starburst  and SN  components  were then determined  based on the starburst archival magnitudes and on the template-subtracted SN photometry simultaneous with the spectra, respectively.    Finally, the template was reduced to the SN redshift and subtracted from the SN spectra in rest-frame.
With this procedure the spectra show some variation in the residual continuum of the blue spectral region, which we attribute to uncertainties in the flux calibration. We allowed for a small adjustment in the template continuum slope (corresponding to  $\pm$0.1 mag variation in $E_{B-V}$)  to ensure  all spectra show a similar overall continuum.

% -----------  Table 2: Log of ground-based spectroscopic observations  --------------
\begin{table*}
\centering
\caption{Ground-based spectroscopy of iPTF15dld.}
\begin{tabular}{cccccc}
\hline
\hline
MJD          &  UT                    &   Phase$^a$        & Telescope  &  Instrument  & grism \\
\hline
57330       &  2015 Nov 4       &  19.1      & 1.82m       &   AFOSC       &     gm4       \\   
57332       &  2015 Nov 6        &  21.0    & TNG          &   LRS            &    LRS-B  \\    
57332       &  2015  Nov 6       &  21.0    & LT              &  SPRAT        &    red        \\
57333       & 2015  Nov 7        &  22.0    &  NTT          &   EFOSC2    &    gr13      \\
57342       &  2015 Nov 16      &  30.6    & 1.82m        &   AFOSC      &   gm4        \\
57344       &  2015 Nov 18      &  32.5    & 1.82m        &   AFOSC      &   gm4        \\
57360       &  2015  Dec 4       &  47.8    &  NTT          &   EFOSC2    &     gr13      \\
57373       &  2015  Dec 17     &  60.2    &  LT              &  SPRAT       &    red        \\
57374       &  2015  Dec 18     &  61.2   &  LT              &  SPRAT       &    red        \\
\hline
\hline 
\noalign{\smallskip}
\multicolumn{6}{l}{$^a$  Phase is given in days with respect to light curve maximum and in rest frame.}\\
\end{tabular}
\label{tab:gb_obs_spec}

\end{table*}
% -----------------------     END OF TABLE 2    ---------------------------

% -------------------   Table 3: Log of Swift/UVOT observations  -----------------

\begin{table*}
\centering
\caption{{\it Swift}/UVOT observations of the region of iPTF15dld on 6-7 November 2015$^a$.}
\begin{tabular}{cccc}
\hline
\hline
Filter & Expotime (s) & Vega mag$^b$  & AB mag$^b$ \\
\hline
$v$      &  508.36    &  17.82 $\pm$ 0.08 (stat) $\pm$  0.01 (sys)  & 17.81 $\pm$ 0.08 (stat) $\pm$ 0.01 (sys)  \\
$b$      &   706.64   &  18.24 $\pm$ 0.05 (stat)  $\pm$ 0.02 (sys)   &  18.12 $\pm$  0.05 (stat) $\pm$  0.02 (sys)  \\
$u$      &   706.65   & 17.72  $\pm$ 0.05 (stat)  $\pm$ 0.02 (sys)   & 18.74  $\pm$ 0.05 (stat) $\pm$ 0.02 (sys)  \\
$uvw1$   &  1415.24   & 17.55 $\pm$ 0.04 (stat) $\pm$  0.03 (sys)   &  19.08 $\pm$ 0.04 (stat) $\pm$ 0.03 (sys)  \\
$uvm2$   &   2576.14  & 17.51  $\pm$ 0.03 (stat) $\pm$ 0.03 (sys)   & 19.10 $\pm$ 0.03 (stat) $\pm$ 0.03 (sys)  \\
$uvw2$   &   2576.14  & 17.51  $\pm$ 0.03 (stat) $\pm$  0.03 (sys)  &  19.20 $\pm$ 0.03 (stat) $\pm$ 0.03 (sys)  \\
\hline
\hline 
\noalign{\smallskip}
\multicolumn{4}{l}{$^a$  Note that these measurements refer entirely to the emission of the starburst region underlying} \\  
\multicolumn{4}{l}{ ~ ~ the SN, while  the SN itself is undetected at these wavelengths.} \\
\multicolumn{4}{l}{$^b$  Not corrected for Galactic extinction.} \\
\end{tabular}
\label{tab:uvotobs}

\end{table*}

% -----------------------------------     END OF TABLE 3   --------------------------------------

%%%%%%%     SECTION 3: RESULTS    %%%%%%%

\section{Results}

\subsection{Host galaxy}
\label{res:hostgalaxy}

The SN is hosted by a compact starburst galaxy/region that, in turn, appears projected over the disc of a 
spiral galaxy.    The   narrow emission lines we detected in our spectra (see Sect. \ref{res:spectra}) indicate that the two objects, starburst and spiral  galaxy,  are located at the same redshift, although we cannot assess whether they form a unique structure or a galaxy pair.   The starburst nucleus is a luminous UV source which  was detected by GALEX on 8 October 2008  (GALEX source J005813.0-033946) with AB magnitudes FUV = 18.89, and NUV = 18.38, (Kron aperture; note that the NUV band, $\sim$2300 \AA,  is similar to the uvm2 band of {\it Swift}/UVOT).

The  SDSS magnitudes of the starburst region at the location of the SN are reported in Table \ref{tab:hostgalaxy}.  Note  that the half-magnitude offset in the measurements obtained with different  photometric apertures  does not affect significantly the colours.  
The $u$-band  magnitude obtained with the $5^{{\prime}{\prime}}$ radius aperture, $u = 19.1$ mag, is consistent with the AB magnitude measured by UVOT in the U-band (Table \ref{tab:uvotobs}).  This and the lack of UV flux variability  suggest  that  the source detected by UVOT is  dominated by the emission of the starburst region, so that the UV emission of the SN is undetectable. 
At a distance of 200 Mpc, the starburst component has an absolute magnitude in $g$-band  of -18.5 mag, which places it at the bright end of the blue compact dwarf luminosity function \citep{Tolstoy2009}.

\begin{figure}
\includegraphics[width=1.0\columnwidth]{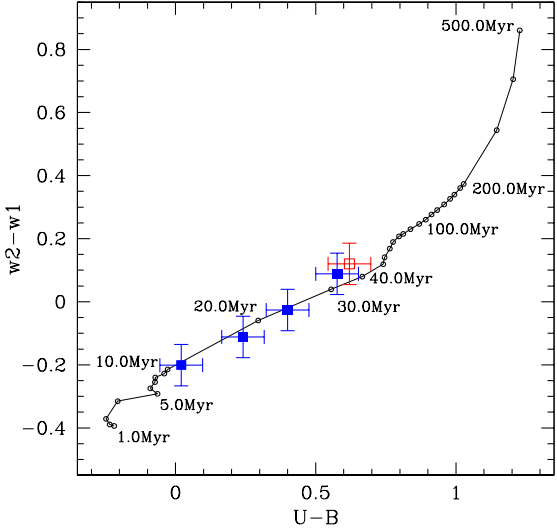}
\caption{Stellar synthesis diagram for the starburst region underlying  iPTF15dld.   Ages of the stellar populations along the diagram are indicated.   The squares represent the observed (empty red) and de-reddened (filled blue) colours of the starburst, obtained from the magnitudes reported in Table \ref{tab:uvotobs} by correcting for different amounts of internal absorption ($E_{B-V}$ = 0.027, 0.137, 0.237, 0.377 mag) and  using the circumstellar Large Magellanic Cloud extinction law with no red-tail-corrected coefficients of Brown et al. (2010; see their Table 1).
For maximum extinction ($E_{B-V}$ = 0.377 mag), the starburst is compatible with an age of 10 Myr, equivalent to the lifetime of a 20 M$_\odot$ star.}
\label{fig:hostgalpopsynth}
\end{figure}

Figure \ref{fig:hostgalpopsynth} shows a stellar population synthesis model to estimate the age of the stellar 
population in the vicinity of the SN from the observed colours \citep{Brocato2000,Raimondo2009}.  The model 
assumes solar metallicity and ages comprised between 1 and 500 Myr.    By correcting the starburst colours -- 
adopting the circumstellar Large Magellanic Cloud extinction law of \citet{Goobar2008} as in \citet{Brown2010} --
for moderate values of intrinsic extinction (from null to $E_{B-V} = 0.35$, i.e. somewhat higher than the maximum intrinsic extinction of the assumed star-forming galaxy template, $E_{B-V} = 0.21$),  in addition to the Galactic one ($E_{B-V} = 0.027$),  we obtain the intrinsic colours reported in Figure \ref{fig:hostgalpopsynth} as filled blue squares.  The colour resulting from maximum correction is consistent with a population age of 10 Myr, which corresponds to the evolution time of a 20 M$_\odot$ star.  The use of an extinction curve more suitable for hot stars \citep{Siegel2014} leads to a similar conclusion.  

This satisfactory match indicates the presence of a young massive star population, consistent with the explosion of a massive stellar core that has evolved from a main sequence mass of $\sim$20 M$_\odot$ (see Section \ref{sect:discussion}).      We note that a Milky Way extinction curve only provides a match with the starburst colors if the intrinsic extinction is as high as $E_{B-V}$ = 0.8, which is inconsistent with the observed colours of the starburst and indicates that this  region presents the characteristics of a more rapidly star-forming,  lower metallicity, less evolved environment than our Galaxy.
In fact, the star-formation rate of $\sim$1~M$_\odot$~yr$^{-1}$ derived by \citet{Palli2016} from radio excess detection within a region a few kpc across, spatially compatible with the UVOT source, points to an explosion site of high star formation rate per unit mass.  This is typical for stripped-envelope SNe \citep{Anderson2012,Crowther2013}, 
expected to be predominantly associated with bright   regions of massive and rapid star formation, which could make their detection systematically more arduous at large distances even with the biggest telescopes.

% ---------------------    Table 4:  Starburst magnitudes    --------------------------

\begin{table}
\centering
\caption{Magnitudes$^a$ of the starburst region.}
\begin{tabular}{ccc}
\hline
\hline
Filter$^a$ &  $5^{{\prime}{\prime}}$-radius &  $3^{{\prime}{\prime}}$-radius  \\
\hline
$u$   &     19.09  & 19.60    \\
$g$   &     18.03  & 18.59  \\
$r$     &   17.77   &  18.43   \\ 
$i$    &   17.50  & 18.26    \\ 
$z$    &    17.46  & 18.22   \\
\hline
\hline 
\noalign{\smallskip}
\multicolumn{3}{l}{$^a$  in the SDSS system, not corrected}\\
\multicolumn{3}{l}{ ~~   for Galactic extinction.} \\
\end{tabular}
\label{tab:hostgalaxy}
\end{table}

% ----------------------------     END OF TABLE 4    -------------------------

%------------- Figure 2:  Population Synthesis for host galaxy --------------
%----------------------------------------------------------------------------------------------------

\subsection{Light curves}
\label{res:LCs}

\begin{figure}
\includegraphics[width=1.0\columnwidth]{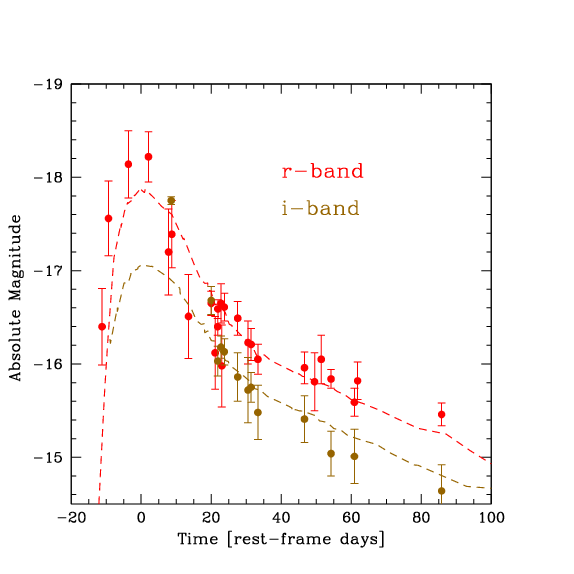}
\caption{Light curves of  iPTF15dld in $r$-band  (red circles) and $i$-band (brown circles), corrected for 
Galactic extinction ($A_V = 0.085$).    At $z = 0.047$, the central wavelengths of these bands correspond to 5980 \AA\ and 7328 \AA, respectively.  The time origin corresponds to the maximum of the $r$-band light curve.    For comparison, we overlaid the light curves of the type Ic SN~2007gr at identical reference wavelengths (dashed curves; see text for the construction of these templates).  The ``$r$-band"-equivalent template of  SN~2007gr was brightened by 0.7 magnitudes for best match with iPTF15dld.}
\label{fig:snlcs}
\end{figure}

The $r$- and $i$-band magnitudes  of the point-like SN source, derived with PSF fitting from the background-subtracted images (see Section \ref{data:photred}),  are reported in Table~\ref{tab:gb_obs_phot} and, after correction for Galactic absorption (using $A_V = 0.085$, Schlafly \& Finkbeiner 2011, and the extinction curve of Cardelli, Clayton \& Mathis 1989),  in Figure~\ref{fig:snlcs}.     We have not corrected for intrinsic extinction within the starburst region because we cannot estimate how much this influences the SN emission (it depends on the relative position of the SN and starburst with respect to the observer) and we have no evidence that iPTF15dld is significantly absorbed in its rest-frame.  In fact, its $R - I$ color, computed from the $r$- and $i$-band light curves,  is comparable to that of well-monitored SNe Ic close to maximum luminosity  \citep{Richmond1996,Galama1998,Patat2001,Foley2003,Ferrero2006,Taubenberger2006,Valenti2008a,Valenti2008b,Hunter2009}, and possibly bluer at later times,  likely owing to significant background still affecting the weaker $r$-band flux.
No detection of iPTF15dld was obtained with the $ugz$ filters in individual exposures.  The magnitudes from the co-added exposures in these filters are consistent with the SDSS measurements.

The $r$- and $i$-band light curves of iPTF15dld were compared with those of SN~2007gr, a type Ic SN of  ``classical" spectral appearance, i.e. with no broad absorption lines \citep{Valenti2008b,Hunter2009}.    At $z = 0.047$, the central wavelengths of the $r$- and $i$-band filters  correspond to 5980 \AA\  and 7328 \AA, respectively.  From the $VRI$ light curves of SN~2007gr we have constructed template light curves at those two reference wavelengths and reported them in Figure \ref{fig:snlcs}, after brightening  the template at 5980 \AA\ by 0.7 magnitudes.  With the exception    of the first $i$-band point, which is significantly brighter, the match with the templates is generally satisfactory, and it indicates that iPTF15dld is a factor of $\sim$2 brighter at $\sim$6000 \AA\ and therefore
bluer  than SN~2007gr in the 6000-8000 \AA\ range.

\begin{figure*}
\includegraphics[width=0.9\textwidth]{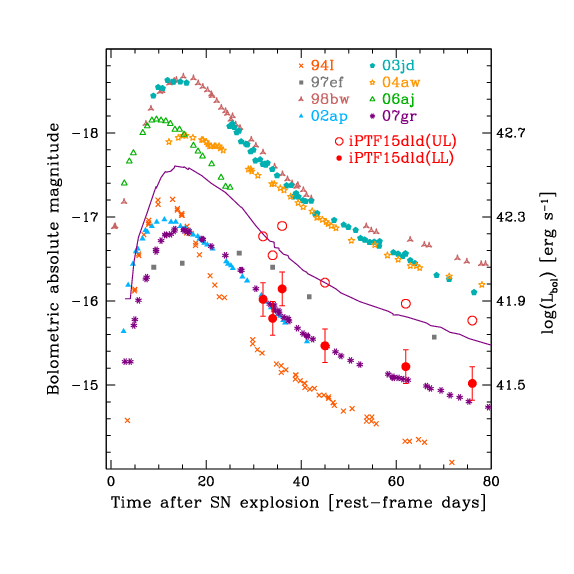}
\caption{Pseudo-bolometric (UVOIR) light curves of stripped-envelope SNe.  The curve of iPTF15dld was obtained by integrating the spectral flux in its rest-frame (filled red points).  Since this covers a limited wavelength range ($\sim$3800-7800 \AA), it is likely a lower limit (LL) on the UVOIR light curve, and  a correction of a factor of 2 was applied to take into account the flux in a broader range (3300-24000 \AA), based on the ratio of broad-band optical and near-infrared fluxes in SNe 1998bw, 2004aw, 2007gr.  These corrected pseudo-bolometric luminosities, that can be considered an upper limit (UL) on the UVOIR light curve, are reported as open red circles.  
The errors on the iPTF15dld luminosities are estimated to be $\sim$20\%. For clarity, the errors on the bolometric luminosities of all other SNe were omitted (the data for these are from Iwamoto et al. 2000; Ferrero et al. 2006; Hunter et al. 2009, and references therein; the data of SN~1997ef were corrected for the different value of the Hubble constant adopted here).  The purple curve represents the bolometric light curve of SN~2007gr brightened by 0.75 mags.}
\label{fig:bolsnlcs}
\end{figure*}

Although  the available photometry ($r$- and $i$-band only) is not sufficient to  construct a proper pseudo-bolometric light curve, the total spectral flux is a rough proxy of the  bolometric behavior.  For each spectrum, we   integrated the flux-calibrated, dereddened  spectral signal in the rest-frame, approximately corresponding to the range 3800-7800 \AA\  (see Fig. \ref{fig:spectra}) and  obtained a bolometric light curve that is similar in shape  to those of the faintest  stripped-envelope SNe  that were monitored long enough to allow a comparison with iPTF15dld (SNe 1994I, 2002ap) and in particular to that of SN~2007gr (see Hunter et al. 2009).  Since our pseudo-bolometric estimate does not include the near-UV and near-infrared contributions, we have estimated this using other SNe Ic that have good photometric coverage in these bands simultaneous with the optical.   At epochs comparable to those of the iPTF15dld photometry, the near-UV and near-infrared fluxes of type  SNe Ic combined represent about 40-50\% of the total flux in 3000-24000 \AA\  (e.g. SN~1998bw, Patat et al. 2001; SN~2004aw, Taubenberger et al. 2006; SN~2007gr, Hunter et al. 2009).   Even taking this into account, iPTF15dld is still  less luminous than the average of stripped-envelope SNe (Fig. \ref{fig:bolsnlcs}).

%------------ Figure 3:  Monochromatic Light curve(s)  ---------------------------
%-----------------------------------------------------------------------------------------------

\subsection{Spectra}
\label{res:spectra}

\begin{figure}
\includegraphics[width=1.5\columnwidth]{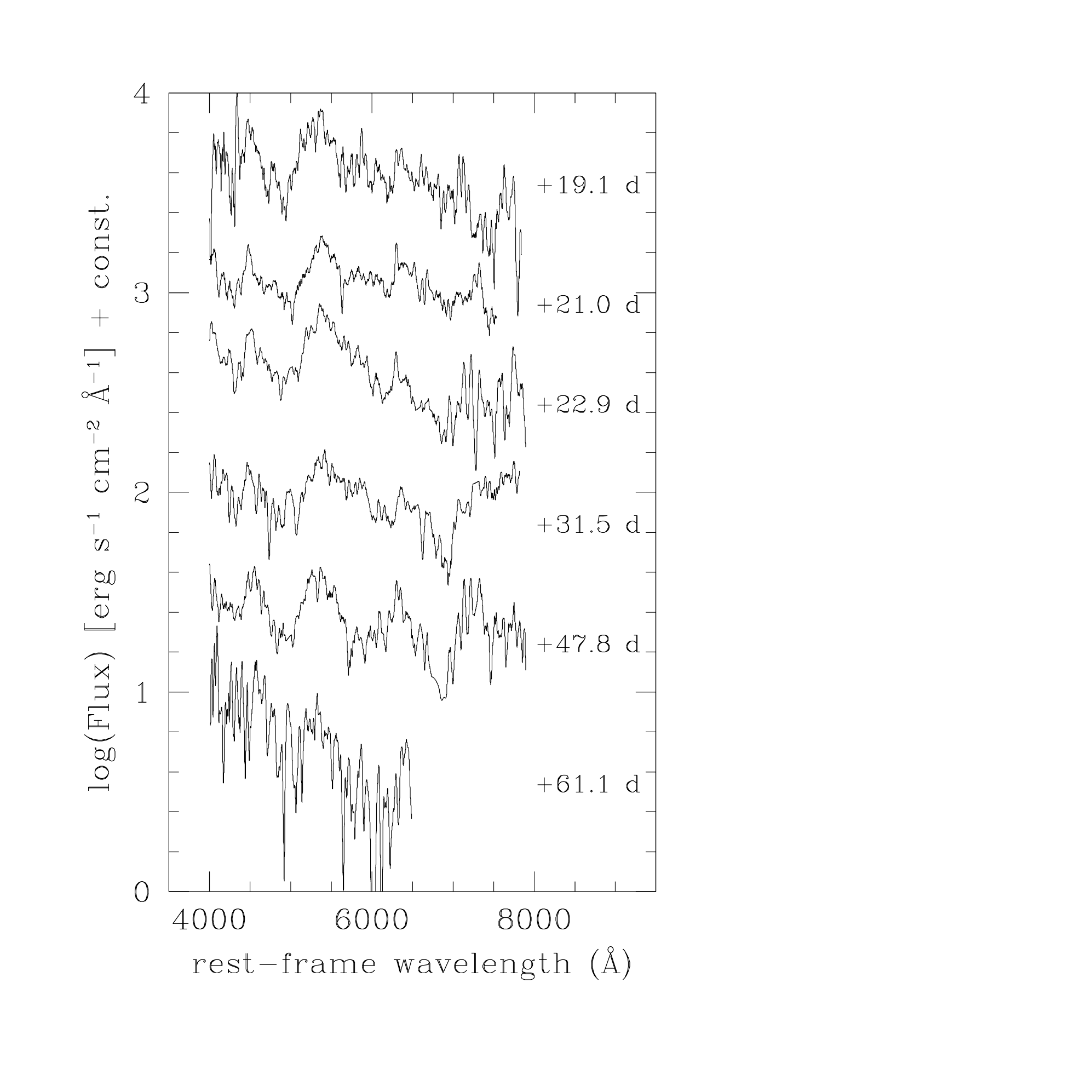}
\caption{Spectra of iPTF15dld in rest frame, corrected for Galactic extinction ($A_V = 0.085$),  smoothed with a boxcar of 50 \AA\ and arbitrarily scaled in flux.  The phases are given in rest-frame, with respect to maximum luminosity.}
\label{fig:spectra}
\end{figure}

The two spectra taken  at the 1.82m {\it Copernico} telescope on 16 and 18 November 2015 were averaged, owing to their closeness in time and similarity, and so were the two spectra acquired at the LT with SPRAT on 17 and 18 December 2015.  
Six final spectra, corrected for Galactic extinction and redshift,  are reported in Figure ~\ref{fig:spectra}.    The SPRAT spectrum of November 6 was not shown because it is very close in time to the TNG spectrum and of lower signal-to-noise ratio.  
The starburst dominates the spectral emission with a blue continuum and narrow emission lines.  
However, when its contribution is removed (see Section \ref{data:specred}), the broad lines typical of  SNe Ic become visible in the visual/red spectral regions.
No hydrogen nor helium absorption lines are seen, indicating a  high degree of envelope stripping and leading to type Ic classification of the SN. The narrow emission lines from the underlying starburst region  were removed.

In search of a close spectral analogue of  iPTF15dld, we compared its  spectra with those of eight type Ic SNe,  both broad- and narrow-lined (SN~1994I, Filippenko et al. 1995; Richmond et al. 1996; Millard et al. 1999; SN~1997ef, Iwamoto et al. 2000; Mazzali et al. 2000; SN~1998bw, Patat et al. 2001; SN~2002ap, Gal-Yam et al. 2002; Mazzali et al. 2002; Foley et al. 2003; SN~2003jd, Valenti et al. 2008a; SN~2004aw, Taubenberger et al. 2006; SN~2006aj, Mazzali et al. 2006; SN~2007gr, Hunter et al. 2009).   
With the partial aid of  a $\chi^2$-minimization routine we selected the spectra of our  SN templates that best-matched, in the 4000-7500 \AA\ wavelength range,  those of iPTF15dld at comparable phases after  light curve maximum.   

\begin{figure}
\includegraphics[width=1.0\columnwidth]{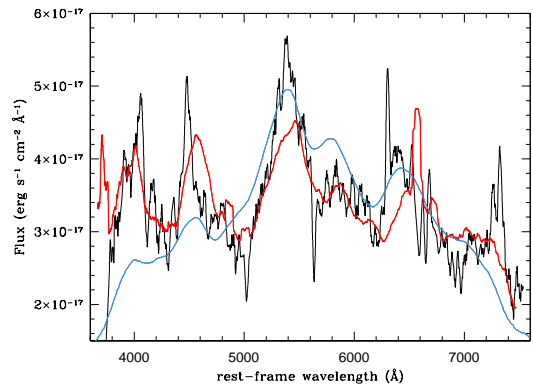}
\caption{Spectrum of iPTF15dld of 6 November 2015 (black) dereddened with $A_V = 0.085$   compared with those of  SN~1997ef (red) and SN~1998bw (blue) at comparable rest-frame phases. The spectrum of SN~1998bw was
dereddened with $A_V = 0.16$, while that of SN~1997ef needs no absorption correction.
All spectra were smoothed with a boxcar of 50 \AA.  The absorption  lines of SN~1998bw are significantly broader than those of iPTF15dld, while those of SN~1997ef represent a better match.}
\label{fig:psc1}
\end{figure}

\begin{figure}
\includegraphics[width=1.0\columnwidth]{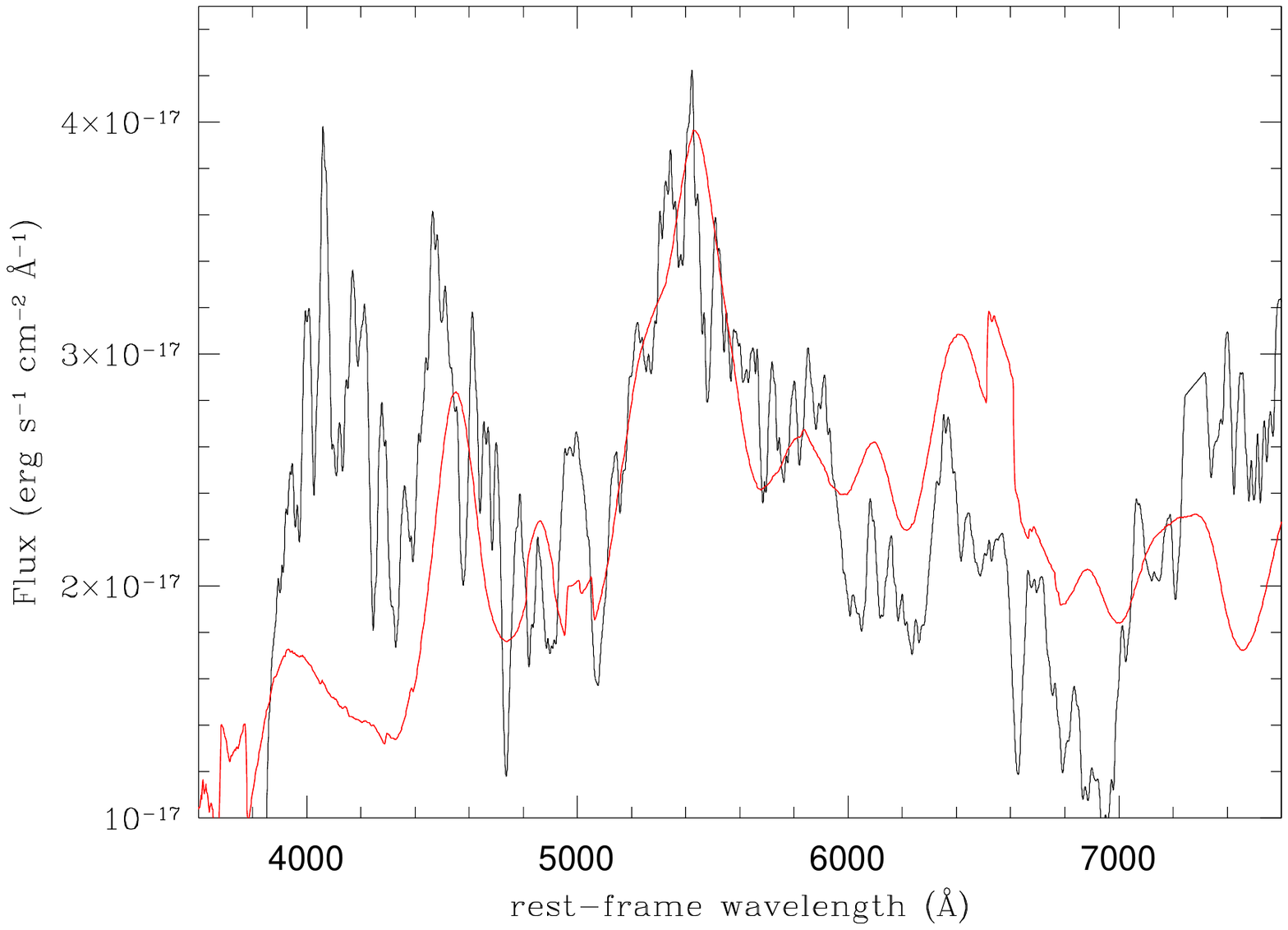}
\caption{Spectrum of iPTF15dld of  17 November 2015 (black) dereddened with $A_V = 0.085$  compared with that of  SN~2003jd (red) at comparable rest-frame phase,  dereddened with $A_V = 0.43$.
All spectra were smoothed with a boxcar of 50 \AA.}
\label{fig:psc2}
\end{figure}

SNe 1998bw and 2006aj, that were associated with GRBs \citep{Galama1998,Pian2000,Pian2006,Campana2006}, do not  compare well with iPTF15dld because their spectra have  significantly broader absorption lines  (although in the case of SN2006aj only one spectrum overlaps in phase).  On the other hand, the classical SNe 1994I and 2007gr represent an equally unsatisfactory match because they have  narrower lines than  our target. 
The first four spectra of iPTF15dld are more similar to those of SNe 1997ef, 2002ap, 2003jd and 2004aw, that are  
broad-lined Ic SNe with no accompanying GRB  (see also Corsi et al. 2016).  These have kinetic energies higher than seen on average in SNe Ic, although they are not as massive nor as luminous as GRB SNe.    The last spectra (December 2015) resemble both  broad- and narrow-lined Ic SN spectra, presumably because they are more noisy and at those epochs ($\sim$50-60 rest-frame days after maximum), the photospheric velocities have significantly decreased also in broad-lined SNe.  In Fig. \ref{fig:psc1} and \ref{fig:psc2} we show  two examples of spectral comparison.

While the signal-to-noise ratio of the spectra and the partial blending of absorption lines, due to their width,  makes it difficult to isolate the chemical species and measure their associated velocities, the similarity with broad-lined SNe suggests higher-than-normal  photospheric velocities.

%------------------ Figure 4:   Bolometric Light curve(s)  --------------------
%-------------------------------------------------------------------------------------------

%--------------------- Figure 5:  spectral sequence  ------------------------------
%----------------------------------------------------------------------------------------------------

%---  Figure 6:  spectral comparison of Sp.2 with SN1997ef  and SN1998bw   ---
%-----------------------------------------------------------------------------------------------

%----------- Figure 7:  spectral comparison of Sp.4 with SN2003jd   ---------
%----------------------------------------------------------------------------------------------

%%%%%%%%     SECTION 4: DISCUSSION    %%%%%%%%%

\section{Discussion}
\label{sect:discussion}

The light curve of  iPTF15dld resembles that of normal, narrow-lined type Ic SNe, with  SN~2007gr (Hunter et al. 2009) providing an excellent match (Fig.~ \ref{fig:snlcs}).  However, the photospheric absorption  lines are broad, so this is classified as a broad-lined Ic SN, rather similar  to well-monitored broad-lined SNe Ic at comparable epochs after  light maximum (SNe 1997ef, 2002ap, 2003jd, 2004aw).
Since spectra were taken only starting 20 days after maximum, we cannot make an assessment of the photospheric velocity before and around maximum;  similarly, the photometric information does not allow us to construct a pseudo-bolometric light curve covering the epoch of maximum luminosity.  As a consequence, our estimates of the physical parameters are only approximated.

In absence of synthetic light curve and spectra based on a detailed radiative transfer model obtained from observed quantities, the  basic SN physical parameters can be derived by rescaling those of other well studied SNe using the 
fundamental relationships of \cite{Arnett1982}, as done for instance in  \cite{Corsi2012,Mazzali2013,Walker2014,Delia2015}.  However, iPTF15dld lacks an estimate of both its light curve width, $\tau$, and its photospheric velocity at maximum luminosity, $v_{ph}$.   Therefore, our estimate of its kinetic energy and ejecta mass can only be based on an average of these parameters for the five SNe that provide the best light curve and spectral match (see Section \ref{res:spectra}).  

From the physical parameters estimated for SNe 1997ef, 2002ap, 2003jd, 2004aw  and 2007gr, \citep{Iwamoto2000,Mazzali2000,Mazzali2002,Valenti2008a,Taubenberger2006,Hunter2009} we derive ranges of [1-18] $\times 10^{51}$ erg and  [2-10] M$_\odot$
for the kinetic energy and ejecta mass of iPTF15dld, respectively.  
Since the shape and luminosity of the bolometric light curve suggest that iPTF15dld could have been similar to SN~2007gr or up to a factor of 2 more luminous at peak, we accordingly estimate that the mass of radioactive $^{56}$Ni synthesized  in the explosion may be in the interval [0.08-0.2] M$_\odot$.
These values are consistent with a progenitor of main sequence mass of the order of $\sim$20-25 M$_\odot$.  A dedicated accurate model is not completely justified by the limited quality of these data.   

Broad-lined Ic SNe of modest luminosity are a rather uncommon and poorly known class, and have started to be detected in larger numbers thanks to dedicated surveys.   As GRB SNe, that are significantly more massive and luminous,   they may be partially powered by an inner engine, i.e. an unusual type of remnant, like a magnetar or a black hole.   The prototype of this sub-class is SN~2002ap \citep{Mazzali2002}, for which evidence had been found of a small fraction of ejected material accelerated to velocities larger than 30000 km~s$^{-1}$.   Since these objects have low ejecta mass (their synthesized $^{56}$Ni mass is small), the total kinetic energy is also not extremely large ($\sim 10^{51}$ erg), but the high photospheric velocities suggest a powerful engine.   Whether these are the progenitors of GRBs that are misaligned with respect to the line of sight and therefore go undetected, or they represent a population of intermediate properties between classical, narrow-lined SNe Ic and  GRB SNe, is matter of controversy  \citep{Mazzali2005,Maeda2008,Soderberg2010,Pignata2011}.  Clarification of this issue  \citep[e.g., through late-epoch radio observations, ][]{vanEerten2011} may lead to a simplification of the apparent  diversity of  stripped-envelope SNe.
We note that the opposite, i.e. low photospheric velocities in highly luminous SNe are never observed   \citep[e.g., ][]{Mazzali2013}.

The case of iPTF15dld shows how optical surveys that cover large areas of the sky  with good cadence using classical facilities   can improve dramatically the study of a broad range of transients. 
Early detection and decent monitoring of objects with a variety of properties will fill  gaps present in the current  information and unify seemingly different  phenomena.

\section*{Acknowledgements}

We acknowledge data from the Pan-STARRS1 telescope which is  supported by NASA under grant No. NNX12AR65G and grant No. NNX14AM74G issued through the NEO Observation Program.  We acknowledge funding from ASI INAF grant I/088/06/0, from the Italian Ministry of Education and Research and the Scuola Normale Superiore,  and from INAF project: ``Gravitational Wave Astronomy with the first detections of aLIGO and aVIRGO experiments" (P.I.: E. Brocato). 
 GR is partially supported by the PRIN INAF-2014 ``EXCALIBURS: EXtragalactic distance scale CALIBration Using first - Rank Standard candles" (PI G. Clementini).
L. Tomasella, EC, and SB are partially supported by the PRIN-INAF 2014 project  ``Transient Universe: unveiling new types of stellar explosions with PESSTO".  MS  gratefully acknowledges generous support provided by the Danish Agency for Science and Technology and Innovation realized through a Sapere Aude Level 2 grant.
MB, GG and GS acknowledge financial support from the Italian Ministry of Education, University and Research (MIUR) through grant FIRB 2012 RBFR12PM1F.
We thank the staff of the {\it Copernico} telescope, LT, NTT, TNG, NOT, and LSQ, in particular W. Boschin, D. Carosati, S. Dalle Ave, L. Di Fabrizio, A. Fiorenzano,  P. Ochner, T. Pursimo and T. Reynolds.  We are grateful to S. Valenti for sending  archival supernova data in digital form,  to P. Nugent  for his support of this project,  to L. Singer for his critical reading of the manuscript,   and to our referee for constructive suggestions.
This research has made use of the Weizmann Interactive Supernova data Repository (http://wiserep.weizmann.ac.il), and it used resources of the National Energy Research Scientific Computing Center, a DOE Office of Science User Facility supported by the Office of Science of the U.S. Department of Energy under Contract No. DE-AC02-05CH11231.

%%%%%%%%%%%%%%%%%%%%%%%%%%%%%%%%%%%%%%%%%%%%%%%%%%



%\bibliographystyle{mnras}


%%%%%%%%%%%%%%%%%%%%%%%%%%%%%%%%%%%%%%%%%%%%%%%%%%

%%%%%%%%%%%%%%%%% APPENDICES %%%%%%%%%%%%%%%%%%%%%

%\appendix


%%%%%%%%%%%%%%%%%%%%%%%%%%%%%%%%%%%%%%%%%%%%%%%%%%

% Don't change these lines
\bsp	% typesetting comment
\label{lastpage}
\end{document}